\documentstyle[twocolumn,prb,aps]{revtex}
\def\revtex{R\raise2pt\hbox{E}VT\lower2pt\hbox{E}X}

  \let\d=\delta
 \let\z=\zeta 
  \let\m=\mu \let\n=\nu
 \let\p=\pi \let\r=\rho \let\s=\sigma 
 \let\f=\varphi  
  \let\D=\Delta

\def\V#1{{\,\underline#1\,}}
\let\dpr=\partial \let\io=\infty \let\ig=\int
\def\fra#1#2{{#1\over#2}}\def\media#1{\langle{#1}\rangle}
\let\0=\noindent
\def\tende#1{\ \vtop{\ialign{##\crcr\rightarrowfill\crcr
\noalign{\kern-1pt\nointerlineskip} \hglue3.pt${\scriptstyle%
#1}$\hglue3.pt\crcr}}\,} \def\otto{\
{\kern-1.truept\leftarrow\kern-5.truept\to\kern-1.truept}\ }
\def\*{\vskip0.3truecm}

\def\lis#1{{\overline #1}}
 \def\ie{\hbox{\it i.e.\ }}

\def\={{ \; \equiv \; }}
\def\defi{\,{\buildrel def \over =}\,}

\def\W#1{#1_{\kern-3pt\lower6.6truept\hbox to 1.1truemm
{$\widetilde{}$\hfill}}\kern0pt}

\def\uu{{\V u}}\def\kk{{\V k}}\def\xx{{\V x}}
\def\cfr{{\it cf.\ }}\def\DD{{\cal D}}
\def\*{\vskip0.3cm}

\def\FINE{
\0{\it
Preprints at:
{\tt http://ipparco.roma1.infn.it}\\
\0\sl e-mail: {\tt giovanni.gallavotti@roma1.infn.it}
}}

\def\Eq(#1){\label{{#1}}}\def\eq(#1){\label{{#1}}}
\def\equ(#1){(\ref{{#1}})}

\title{Entropy driven intermitency}

\author{G. Gallavotti\\
Fisica, Universit\`a di Roma La Sapienza,\\
P.le Moro 2, 00185 Roma, Italia}

\begin{document}
\maketitle
\begin{abstract}
This note reviews some physical aspects of the chaotic
hypothesis in nonequilibrium statistical mechanics and attempts at
the physical interpretation of the fluctu\-ation theorem as a
quantitative intermittency property.
\end{abstract}
\*\*

\0The main assumption, for the foundation of a nonequilibrium
statistical mechanics or a theory of developed turbulence, will be:
\*

\0{\bf Chaotic hypothesis: \it Asymptotic motions of a chaotic system,
be it a system of $N$ particles or a viscous fluid,
can be regarded as motions of a mixing Anosov system, for the
purposes of computing time averages.}
\*

The point of view goes back to Ruelle, \cite{[Ru78a]}, \cite{[Ru78b]},
and in this specific form has been proposed in \cite{[GC95]}. I shall
not define here ``mixing Anosov system'', see \cite{[Ga99a]}. It
will suffice to say that Anosov systems are very well understood
dynamical systems in spite of being in a sense the most chaotic: they
are so well understood that they can be regarded as the {\it paradigm}
of chaotic systems much as the harmonic oscillators are the paradigm
of regular and orderly motions.

This immediately implies the existence and uniqueness of an invariant
probability distribution $\m$ which gives the {\it statistics}
if the motions of the system in the sense

\begin{equation}
\lim_{T\to\io} T^{-1}\ig_0^T F(S_t x)\,dt=\ig \m(dy)\,F(y)\Eq(1)\end{equation}
for {\it almost all} initial data, \ie outside a set of $0$ volume in
phase space (volume being measured in the ordinary sense of the
Lebesgue measure). Here $S_t$ denotes the time evolution map, solution
of the differential equations of motion, and $\m$ is called the {\it
SRB distribution}.
\*

Particular interest will be reserved to {\it time reversible} systems:
\ie systems for which there is an isometry $I$ of phase space such that

\begin{equation}
I^2=1,\quad and\quad I S_{-t}=S_t I\Eq(2)
\end{equation}

Denoting $\dot x= f(x)$ the equations of motion and $\s(x)$ the
divergence ${-\rm div\,} f(x)=-\sum_j \dpr_{x_j} f_j(x)$ a key quantity
to study will be

\begin{equation}
\s_+={\rm time\ average\ of\ } \s\Eq(3)\end{equation}
that shall be called {\it average entropy production rate} or {\it
average phase space contraction rate}; it will be assumed that
$\s_+>0$ (this quantity is, in general, non negative, \cite{[Ru96]}).

The main result on $\s(x)$ concerns the probability distribution in
the stationary state $\m$ of the quantity (an observable, \ie a
function of the phase space point $x$)

\begin{equation}
p= T^{-1}\ig_{T/2}^{T/2}\fra{\s(S_t x)}{\s_+}\,dt\Eq(4)\end{equation}
which will be called the {\it $T$--average dimensionless entropy
creation rate}. The result is a symmetry property of its probability
distribution $\p_T(p)$: which can be written in the form $\p_T(p)\defi
\,const\, e^{\z(p)T+o(T)}$, defining implicitly $\z(p)$, as it
follows, on general grounds, from the theory of mixing Anosov
systems, \cfr \cite{[Si77]}, see\cite{1} for a more mathematical
statement. The function $\z(p)$ is called the {\it large deviation
rate} for the observable $p$.

Then (see \cite{[Ge98]}: the delicate continuous time extension of a
similar result\cite{[GC95]} for discrete time) \*

\0{\bf Fluctuation theorem: \it The ``rate function'' $\z(p)$ verifies

\begin{equation}
\z(-p)=\z(p)-p\,\s_+\Eq(5)\end{equation}
which is a parameterless relation, valid under the above hypotheses of
chaoticity and of time reversibility. It is a ``mechanical'' identity
valid for systems with arbitrarily many particles (\ie $N=1,2,\ldots
10^{23}, \ldots$).}
\*

Mathematically the above result holds for mixing Anosov flows
which are time reversible. The chaotic hypothesis extends the result
to rather general systems: of course for such systems it is no longer
a theorem much in the same way as the consequences of the ergodic
hypothesis in equilibrium statistical mechanics are not theorems for
most systems to which they are applied.

Therefore we can say that the chaotic hypothesis implies {\it in
equilibrium} (when the equations of motion are Hamiltonian and
therefore $\s(x)\=0$) the ergodic hypothesis (quite clearly): hence it
implies classical statistical mechanics, starting with Boltzmann's
heat theorem which says that $(dU+p dV)/T$ is an exact differential
(with $U,p,V,T$ defined as time averages of suitable mechanical
quantities, see \cite{[Ga99a]}). Likewise, out of equilibrium and in
reversible systems, the chaotic hypothesis implies a general relation
that is given by the fluctuation theorem, valid for systems with
arbitrarily large numbers of particles. Both the heat theorem and the
fluctuation theorem are ``universal'', \ie parameterless, system
independent relations. They could perhaps be considered a curiosity
for $N$ small, but certainly the first, at least, is an important
property for $N=10^{23}$.
\*

It becomes, at this point, clear that one should attempt at an
interpretation of the fluctuation theorem \equ(5). It is however
convenient to analyze it first in more detail to get some
familiarity with the physical questions that it is necessary to
address to grasp its meaning.

We begin with an attempt at justifying the name ``{\it entropy
creation rate}'' for $\s_+$. Without too many comments we quote here the
simple but relevant remark, \cite{[An82]}, that the so called ``{\it Gibbs
entropy}'' of an evolving probability distribution which starts, at
$t=0$, as absolutely continuous with density $\r(x)$ on phase space has
the following property

\begin{equation}
-\fra{d}{dt}\ig \r_t(x)\log \r_t(x)
\,dx=\ig\s(x)\,\r_t(x)\,dx\Eq(6)\end{equation}
where $\r_t(x)=\r(S_{-t}x)\det\fra{\dpr S_{-t}x}{\dpr x}$ is the
evolving phase space density (here ${\dpr S_{-t}x}/{\dpr x}$ is the
matrix of the derivatives of the time evolution map $S_t$). Since the
r.h.s. of \equ(6) formally tends as $t\to\io$ to the average of $\s$
with respect to the distribution $\m$, \ie to $\s_+$, we realize that
\equ(6) is a possible justification of the name used for $\s_+$.
\*

Before proceeding it is also necessary to understand the role of the
reversibility assumption in order to see whether it is a serious
limitation in view of a possible physical interpretation and physical
interest of the fluctuation theorem. Here one can argue that in many
cases an irreversible system is ``{\it equivalent}'' to a reversible
one: in fact several reversibility conjectures have been proposed, see
\cite{[Ga95]}, \cite{[GC95]}, \cite{[Ga96a]} (Sect. 2 and 5),
\cite{[Ga96b]} (Sec. 8), \cite{[Ga97]}, \cite{[Ga99a]} (Sec. 9.11),
\cite{[Ru99b]}.

Rather than trying to be general I shall consider an example, and analyze
a Navier--Stokes fluid, in a periodic container of side $L$, subject to
a force with intensity $F$ and with viscosity $\n$. If $R=F L^3
\n^{-2}$ is the ``{\it Reynolds number}'', $p$ is the pressure field, the
density is $\r=1$, and $\V g$ is a force field of intensity $1$ the
equations are

\begin{eqnarray}
&\dot{{\V u}}=- R\, \W u\cdot\W\dpr \,\V u+\D\,\V u+\V g- \V\dpr p\nonumber\\
&\V\dpr\cdot\V u=0
\Eq(7)\end{eqnarray}
which are {\it irreversible} equations. According to the K41 theory
(Kolmogorov theory, see \cite{[LL71]}) the above equations can be truncated
retaining only a few harmonics of the field $\V u$: \ie replacing $\V
u(\xx)=
\sum_\kk \uu_\kk \, e^{i\kk\cdot\xx}$ by
$\V u(\xx)=
\sum_{|\kk|<K(R)} \uu_\kk \, e^{i\kk\cdot\xx}$ with $K(R)=R^{3/4}$ so
that ``effectively'' (and accepting the K41 theory) the fluid has
$N(R)=O(R^{9/4})$ degrees of freedom.

If $R$ is large (hence the motion is turbulent) the quantity
$\DD=\ig(\W\dpr\,\V u)^2\,d\xx$ fluctuates in time with some average
$\media{\DD}_{\m_{R}}\defi\DD(R)$, where $\m_{R}$ is the
probability distribution giving the statistics of the time averages.

We can also consider the following equations, introduced in \cite{[Ga97]} and
called GNS equations,

\begin{eqnarray}
&\dot{{\V u}}=- R\, \W u\cdot\W\dpr \,\V u+\n(\uu) \,\D\,\V u+\V g-
\V\dpr p
\nonumber\\
&\V\dpr\cdot\V u=0\Eq(8)\end{eqnarray}
where the ``multiplier'' $\n(\uu)$ is so defined that $\DD(\uu)$ is a
constant of motion. An elementary calculation yields
\begin{equation}
\n(\uu)=\fra{\ig_V\big(\V\f\cdot\D\uu-R\D\uu\cdot(\W
u\cdot\W\dpr\uu)\big)\,d\xx} {\ig_V(\D\uu)^2\,d\xx}\Eq(9)\end{equation}
which shows that $\n(\uu)$ is {\it odd} in $\V u$ so that \equ(8) is
reversible if time reversal is defined as $(I\,\uu)(\xx)=-\uu(\xx)$.

Call ``local observable'' any function $\uu\to F(\uu)$ of the velocity
field which depends on $\V u$ only via fionitely many of its Fourier
compopnents $\uu_\kk$; and let $\tilde \m_{\DD,R}$ be the SRB
distribution for \equ(8). Then the following conjecture was proposed
in \cite{[Ga97]}:

\*
\0{\bf Equivalence conjecture: \it If
$F(\uu)$ is a local observable with non zero average then

\begin{equation}
\fra{\m_{R}(F)}{\m_{\DD(R),R}(F)}\tende{R\to\io} 1\Eq(10)\end{equation}
\ie at large Reynolds number the irreversible NS and the reversible 
GNS equations are equivalent.}
\*

Note the analogy between the conjecture and the equivalence property
of statistical ensembles in equilibrium statistical mechanics: here
the coefficient of the Laplacian in \equ(7) (which is $1$ by our
definitions) plays the role of temperature in a canonical ensemble
while the quantity $\DD$ plays the role of the energy in a
microcanonical ensemble: if the energy is suitably tuned the
distributions $\m_R,\m_{\DD(R),R}$ are statistically equivalent as
$R\to\io$ and the Reynolds number $R$ plays the role of the volume and
$R\to\io$ the role of thermodynamic limit.

It is interesting to remark that in nonequilibrium physics the
statistical ensembles may be defined not only by the parameters that
one regards naturally as control parameters (like energy in the
microcanonical ensemble and temperature in the canonical) but {\it also by
the equations of motion} that are used: this is perhaps not so strange
because in equilibrium systems no friction is necessary and the equations
of motion do not suffer from the ambiguity due to the arbitrariness of
the thermostatting mechanisms that remove heat from the system (making
possible the evolution towards a statistically stationary state).

The above conjecture is beginning to be tested with results that are
at least encouraging, see \cite{[RS99]}. Progress in the theory is needed as
experiments on fluctuations of entropy production are already
available and one would like to interpret them theoretically, \cite{[CL98]}.
\*

The second question that one has to clarify preliminarly is that the
``rate function'' $\z(p)$ in \equ(5) should be expected to be
proportional to some macroscopic parameter measuring the size (like
volume or number of degrees of freedom) so that the probability of
observing the value $p$ in a stationary state is $\p_T(p)= {\,const\,}
e^{\z(p)\,T}$ hence it is not observable if $p\ne1$: any attempt at
measuring $p$ will inexorably lead to $p=1$ (given that by our
normalizations the average value of $p$ is $1$).

Therefore one should investigate if, or when, a ``{\it local
version}'' of the above fluctuation theorem holds telling us some
properties, relative to a small volume or to a few degrees of freedom,
which could have fluctuations that are frequent enough to be
observable.

To get some inspiration we consider an analogous problem: suppose that
a high temperature low density gas has density $\r$ and that it
occupies the whole space. Given a volume $V$ we can consider the
observable $p=N_V/\r V$. Then its probability distribution will have
the form

\begin{equation}
\p_V(p)= \,const\, e^{\lis\z(p)\, V}\Eq(11)\end{equation}
where the exponent is affected by an error $O(\dpr V)$ of the size
of the boundary area of $V$ {\it and $\lis\z(p)$ is $V$--independent}.

This shows that density fluctuations that are not observable in
volumes $V$ of macroscopic size do become observable in small enough
volumes when the quantity $(\lis\z(p)-\lis\z(1))\, V$ becomes
reasonably small. Furthermore since $\lis \z(p)$ is essentially
$V$--independent we can infer that the probability of density
fluctuations in a large volume is also measurable, being trivially
related to $\lis\z(p)$ which is visible via the fluctuations in
small volumes.

Coming back to our dynamical questions we can ask whether there is at
least one model for which one can establish a ``local fluctuation
theorem'' in a sense analogous to the above result on density
fluctuations. Indeed there is a class of models that is very
suitable for illustration purposes: these are the {\it chains of
coupled maps}. Although their importance is mainly illustrative, they
clearly show the possibility of local fluctuation theorems:
here it will be enough to refer to the literature, \cite{[Ga99c]},
\cite{[GP99]}.

The conclusion is that if a local fluctuation relation holds then it
becomes possible to perform tests of the fluctuation theorem,
hence of the chaotic hypothesis.  \*

Therefore we can try to attack the main question of interest here,
namely ``which is the physical interpretation'' of the fluctuation
theorem. The key is the following theorem, which is a simple extension
of it and holds under the same hypotheses (\ie chaotic hypothesis and
time reversibility). It can be regarded as an extension of the
Onsager--Machlup theory of fluctuation patterns, \cite{[OM53]}. Let
$F,G$ be time reversal odd observables (for simplicity and to fix the
ideas): $F(Ix)=-F(x), G(IX)=-G(x)$; and let $h,k: [-T/2,T/2]\to R^1$
be two real valued functions or ``{\it patterns}''. We call
$h'(t)=-h(-t)$, $k'(t)=-k(-t)$ the ``{\it time--reversed patterns}''
or ``{\it antipatterns}'' of the patterns $h,k$. If $F(S_tx)=h(t)$ for
$t\in[-T/2,T/2]$ we say that $F$ follows the pattern $h$ around the
reference point $x$ in the time interval $[-T/2,T/2]\defi W_T$. Then,
see \cite{[Ga99b]},
\*

\0{\bf Theorem \it (extension of Onsager--Machlup theory): The
probabilities of the patterns $h,k$ conditioned to a $T$--average
dimensionless entropy production $p$, see \equ(4), denoted
$\p\big(F(S_t\cdot)\big)=h(t), \, t\in W_T\,\big|\,p\big)$ and
$\p\big(G(S_t\cdot)\big)=k(t), \, t\in W_T\,\big|\, p\big)$
respectively verify

\begin{equation}
\fra{\p\big(F(S_t\cdot)=h(t), \,
t\in W_T\,\big|\, p\big)}{\p\big(F(S_t
\cdot)=-h(-t), \,
t\in W_T\,\big|\, -p\big)}= e^{p\s_+ T}\Eq(12)\end{equation}
and (consequently)
\begin{eqnarray}
\fra{\p\big(F(S_t
\cdot)=h(t), \, t\in W_T\,\big|\, p\big)}{\p\big(G(S_t
\cdot)=k(t), \, t\in W_T\,\big|\, p\big)}= \nonumber\\
=\fra{\p\big(F(S_t\cdot)=
-h(-t), \, t\in W_T\,\big|\, -p\big)}{\p\big(G(S_t\cdot)
=-k(-t), \, t\in W_T\,\big|\, -p\big)}\Eq(13)\end{eqnarray}
Hence relative probabilities of patterns in presence of $T$--average
entropy production $p$ are the same as those of the
corresponding antipatterns in presence of the opposite $T$--average
entropy production rate.}
\*

In other words it suffices to change the sign of the entropy
production to reverse the arrow of time. In a reversible system the
quantity $\z(p)$ measures the degree of irreversibility of a motion
observed to have the value $p$ of dimensionless entropy creation rate
during an observation time of size $T$: if we observe patterns over
time intervals of size $T$ then the fraction of such intervals in
which we shall see an entropy production $p$ rather than $1$ (which is
the most probable value) will be

\begin{equation}
 e^{(\z(p)-\z(1))T}\Eq(14)\end{equation}

More generally, in a situation in which a local fluctuation theorem
holds and $\z(p)=V\lis\z(p)$
we can divide the volume occupied by the system into small boxes of
size $V_0$, small enough so that one can observe entropy production
fluctuations within them, and we can divide the time axis into time
intervals of size $T$. Then
\*

\0{\bf Proposition \it (intermittency of fluctuations): The fraction of
time intervals in which we shall observe $p$ in a given box $V_0$ will
be $e^{(\lis\z(p)-\lis\z(1))\, V_0\,T}$ and this same quantity will be
the fraction of boxes $V_0$ where we shall observe, within a
given time interval of size $T$, entropy production $p$}.  \*

Normally we shall see $p=1$ in a fixed box $V_0$ but ``seldom'' we
shall see $p=-1$ and then, by the above extension of the
Onsager--Machlup theory, {\it everything will look wrong}: every
improbable pattern will appear as frequently as we would expect its
(probable) antipattern to appear.  This will last only for a moment
and then things will return normal for a very long time (as the
fraction of times in which this can happen in a given bix is $e^{-\lis
\s_+ V_0 T}$). This is a kind of {\it intermittency} phenomenon.

In fact we see that when a local fluctuation theorem holds we shall see
intermittency, in the form of a reversed time arrow, {\it happening in a
small volume $V_0$ somewhere} in the volume $V$ of the system, provided

\begin{equation}
{V}\,{V_0}^{-1} e^{-\lis\s_+ V_0\,T}\simeq 1\Eq(15)\end{equation}
furthermore there is a simple relation between fraction of volumes
and fraction of times where time reversal occurs: namely they are
equal and directly measured by $e^{-\lis\s_+ V_0\,T}$, \ie by the
average entropy creation rate.
\*

We conclude by noting that the above remarks set up a possible way of
measuring $\s_+$ and of attempting to measure $\z(p)$ in systems for
which it is hard develop a reasonably good numerical simulation and/or
an expression for $\s_+$.
The quantity $\lis\s_+$ can be measured by considering some
``current'' $J$ associated with the system; \ie an observable $J(x)$
which, among other properties, is odd under time reversal.  One then
looks at the observable obtained by averaging this current over a time
$T$ and averaged over a time $T$, namely $J_T(x)\defi
T^{-1}\ig_{-T/2}^{T/2} J(S_tx)dt$ and one measures how often $J_T$
takes a value close to the opposite of its (infinite time) average
value $J_+$, assuming that the latter is $>0$: this should happen with
a frequency $e^{-\lis\s_+\,V_0\,T}$ giving us access to $\lis\s_+$.

Given the special role that entropy generation plays it is very
tempting to think that there might be many currents $J$ associated
with the system: for each of them one could define $p= J_T/J_+$; then
the new quantity $p$ {\it has the same probability distribution as the
variable with the same name that we have associated with the entropy
production}. This is true at least for the special case $p=-1$ as just
noted: if true in general then we could have easily access to the
function $\z(p)$ for several values of $p$. Hence analysizing this
``universality'' property in special models seems to be an interesting
problem.

For a general review on recent developments in non\-equilibrium
statistical mechanics see \cite{[Ru99a]}.
\*

\0{\bf Acknowledgements: \it This is a contribution to the
Proceedings of ``Inhomogeneous random systems'' January 25-26, 2000,
(Universit\'e de Cergy-Pontoise, Paris); supported partially by
``Cofinanziamento 1999''.}

\0\revtex\\
\FINE

\begin{thebibliography}{}

\bibitem{1} This means, precisely, that the probability
$P$ that $p$ is in an interval of size $\d$ around the value $p$ is
such that $\lim_{T\to\io} T^{-1}\log P=
\sup_{p}\z(p)$ with the supremum in the interval $\d$, for all $p\in
(-p_{\max},p_{\max})$ and $-\io$ if $p\not \in
[-p_{\max},p_{\max}]$. However I shall briefly say that the
probability density of the variable $p$ is $\p_T(p)={\,const\,}
e^{\z(p)T}$.

\bibitem{[An82]} Andrej, L.: {\it The rate of entropy change in
non--Hamiltonian systems}, Physics Letters, {\bf 111A}, 45--46, 1982.

\bibitem{[BGG97]} Bonetto, F., Gallavotti, G., Garrido, P: {\it Chaotic
   principle: an experimental test}, Physica D, {\bf 105}, 226--252,
   1997.

\bibitem{[CL98]} Ciliberto, S., Laroche, C.: {\it An experimental
verification of the Gallavotti--Cohen fluctuation theorem}, Journal de
Physique, {\bf8}, 215--222, 1998.


\bibitem{[Ga95]} Gallavotti, G.:
{\it Ergodicity, ensembles, irreversibility in Boltzmann and beyond},
  Journal of Statistical Physics, {\bf 78}, 1571--1589, 1995. And {\it
  Topics in chaotic dynamics}, Lectures at the Granada school,
  ed. Garrido--Marro, Lecture Notes in Physics, Springer Verlag, {\bf
  448}, p. 271--311, 1995.

\bibitem{[Ga96a]} Gallavotti, G.: {\it New methods in nonequilibrium gases and
fluids}, Open Systems and Information Dynamics, Vol. {\bf 6}, 101--136, 1999
(original in chao-dyn \#9610018).

\bibitem{[Ga96b]} Gallavotti, G. {\it Chaotic hypothesis: Onsager reciprocity
and
   fluctuation dissipation theorem},
   Journal of Statistical Phys., {\bf 84}, 899--926, 1996.

\bibitem{[Ga97]} Gallavotti, G.:
{\it Dynamical ensembles equivalence in fluid mechanics},
   Physica D, {\bf 105}, 163--184, 1997.

\bibitem{[Ga99a]} Gallavotti, G.:
{\it Statistical mechanics. A short treatise}, p. 1--345,
  Springer Verglag, 1999.

\bibitem{[Ga99b]} Gallavotti, G.:
{\it Fluctuation patterns and conditional reversibility in
  nonequilibrium systems}, Annales de l' Institut H. Poincar\'e,
  {\bf70}, 429--443, 1999.

\bibitem{[Ga99c]} Gallavotti, G.:
{\it A local fluctuation theorem}, Physica A,
   {\bf 263}, 39--50, 1999. And: {\it Chaotic Hypothesis and
   Universal Large Deviations Properties}, Documenta Mathematica,
   extra volume ICM98, vol. I, p. 205--233, 1998, also in chao-dyn
   9808004.

\bibitem{[GC95]} Gallavotti, G., Cohen, E.G.D.: {\it Dynamical
ensembles in non-equilibrium statistical mechanics}, Physical Review
Letters, {\bf74}, 2694--2697, 1995. Gallavotti, G., Cohen,
E.G.D.: {\it Dynamical ensembles in stationary states}, Journal of
Statistical Physics, {\bf 80}, 931--970, 1995.

\bibitem{[GP99]} Gallavotti, G., Perroni, F.: {\it An experimental test of the
local fluctuation theorem in chains of weakly interacting Anosov
systems}, mp$\_$arc \#99-???.

\bibitem{[Ge98]} Gentile, G.: {\it Large deviation rule for Anosov flows},
Forum Mathematicum, {\bf10}, 89--118, 1998.

\bibitem{[LL71]} Landau, L., Lifchitz, E.: {\sl M\'ecanique
des fluides}, MIR, Mosca, 1971.

\bibitem{[OM53]}
Onsager, L., Machlup, S.: {\it Fluctuations and irreversible
processes}, Physical Review, {\bf91}, 1505--1512, 1953. And Machlup,
S., Onsager, L.: {\it Fluctuations and irreversible processes},
Physical Review, {\bf91}, 1512--1515, 1953.

\bibitem{[RS99]} Rondoni, L., Segre, E.: {\it Fluctuations in two dimensional
reversibly damped turbulence}, Nonlinearity, {\bf12}, 1471--1487, 1999.

\bibitem{[Ru78a]} Ruelle, D.: {\it Sensitive dependence on initial conditions
and turbulent behavior of dynamical systems}, Annals of the New York
Academy of Sciences, {\bf356}, 408--416, 1978.

\bibitem{[Ru78b]} Ruelle, D.: {\it What are the measures describing
turbulence?}, Progress of Theoretical Physics, (Supplement) {\bf64},
339--345, 1978.

\bibitem{[Ru96]} Ruelle, D.: {\it Positivity of entropy production in
nonequilibrium statistical mechanics}, Journal of Statistical Physics,
{\bf 85}, 1--25, 1996. And Ruelle, D.: {\it Entropy production in
nonequilibrium statistical mechanics}, Communications in Mathematical
Physics, {\bf189}, 365--371, 1997.

\bibitem{[Ru99a]} Ruelle, D.: {\it Smooth dynamics and new theoretical
ideas in nonequilibrium statistical mechanics}, Journal of
Statistical Physics, {\bf 95}, 393--468, 1999.

\bibitem{[Ru99b]} Ruelle, D.: {\it A remark on the equivalence of isokinetic
and isoenergetic thermostats in the thermodynamic limit},
IHES 1999, to appear in Journal of Statistical Physics.

\bibitem{[Si77]} Sinai, Y.G.: {\sl Lectures in ergodic
theory}, Lecture notes in Mathematics, Prin\-ce\-ton U. Press,
Princeton, 1977.

\end{thebibliography}
\end{document}